\documentstyle[11pt,epsfig]{article}
\columnwidth\textwidth
\newcommand{\anueR}{\mbox{$\overline{\nu}_{eR}$}}

\newcommand{\nueL}{\mbox{$\nu_{eL}$}}

\newcommand{\anumR}{\mbox{$\overline{\nu}_{\mu R}$}}

\newcommand{\peae}{\mbox{$P(\nueL\to\anueR)$}}
\newcommand{\dms}{\mbox{$\Delta m^2$}}
\newcommand{\avrho}{\mbox{$\langle \rho\rangle$}}
\newcommand{\avrhot}{\mbox{$\langle \rho(t)\rangle$}}
\newcommand{\avrhop}{\mbox{$\langle\rho'\rangle$}}
\newcommand{\avrhopt}{\mbox{$\langle\rho'(t)\rangle$}}

\begin{document}
\begin{flushright}
FTUV/IFIC-9833\\
hep-ph/9807426.
\end{flushright}

\begin{center}
\Large{\bf Solar Antineutrinos from Fluctuating 
Magnetic Fields at Kamiokande.}
\end{center}

\begin{center}
E. Torrente-Lujan.\\
{\em
IFIC-Dpto. Fisica Teorica. CSIC-Universitat de Valencia. \\
Dr. Moliner 50, E-46100, Burjassot, Valencia, Spain.\\
e-mail: e.torrente@cern.ch}
\end{center}

\begin{abstract}
We consider the effect of a 
strongly chaotic magnetic field at the narrow bottom 
of the convective zone of the Sun  together with 
 resonant matter oscillations on the production of electron  Majorana 
antineutrinos. 
Even for moderate levels of 
noise, we show that it is possible to obtain a small but significant 
probability for    $\nu_e\to \overline{\nu}_e$ conversions (1-3\%)  at 
the energy range 2-10 MeV  for large regions of the mixing parameter space 
while still satisfying present  (Super)-Kamiokande antineutrino 
bounds and observed total rates.
In the other hand it would be  possible to obtain information about the 
solar magnetic internal field if 
antineutrino bounds reach the $1\%$ level and a particle physics 
solution to the SNP is assumed.  
The mechanism presented here has the advantage of being 
  independent of the largely unknown  magnetic profile of the Sun 
and the intrinsic  neutrino magnetic moment.\\

\end{abstract}



\vspace{0.5cm}
\newpage

{\bf 1.} The combined action of  spin flavour precession in a magnetic field
 and  ordinary 
neutrino matter oscillations can produce an observable 
flux of $\anueR$'s from the Sun 
in the case of the neutrino being  a Majorana particle.

 Water Cerenkov detectors such as Kamiokande and
 SuperKamiokande  (SK) which are 
sensitive to the $\nu_e-e$ elastic interaction are 
also capable of detecting these 
$\overline{\nu}_e$'s coming from the sun. Forthcoming solar 
neutrino experiments, such 
as SNO and Borexino are also expected to have  a  high 
sensitivity to them.

The specific signature of electron antineutrinos in proton
 containing materials is the inverse beta decay process:
 $\overline{\nu}_e+p\to n+e^{+}$, which produces 
almost isotropical   monoenergetic positrons with a relatively high 
cross section. Antineutrino events would contribute in this way to the 
isotropic background.

Kamiokande data yielded restrictive limits on solar \anueR\ in  the 
high energy region: for $E_\nu>9.3$ MeV the antineutrino total 
flux does not exceed $6-10\%$ of the SM predictions for the $\nu_e$  
flux in the same energy range \cite{akh6,semi2}.
Fiorentini et Al. \cite{fio1} have shown  that a better 
sensitivity can be achieved by exploiting the huge statistics of SK in 
conjunction with the residual directionality of the positrons from inverse beta 
decay. They pointed out that the angular dependence of the cross section 
can be 
used as a "signature" for the presence of positrons from antineutrinos. Using 
their method not only upper bounds can be obtained but antineutrinos can actually be  detected. 
From  SK data corresponding to the first 101.9 days of 
observation they obtain a model independent bound
$\Phi_{\overline{\nu}}(E_{\overline{\nu}}>8.3$ MeV$)<9.10^4$ cm$^{-2}$
 s$^{-1}$  (C.L. 95\%).
This bound corresponds approximately to a fraction $\approx 5.2\%$ of the solar 
neutrino flux in the energy range considered predicted by the SSM. 
It is claimed that in a 3 year period of data taking SK could be sensitive to 
$\nu_e-\overline{\nu}_e$ oscillations at the $1-3\%$ level.

Bounds on the fraction of $\overline{\nu_e}$ can be translated into 
constraints on the neutrino 
mixing parameters if a specific model is assumed. 
Alternatively, solar properties associated 
with specific astrophysical 
models (i.e. convective solar magnetic fields) can be 
restricted by the absence of observance of antineutrinos.
In the positive side, if we search for antineutrino appearance,  
the main question is to identify  possible
mechanisms which, over large regions of parameter space 
(i.e. without excessive fine-tuning) produce simultaneously relatively small quantities of 
$\overline{\nu_e}$  at Kamiokande energies in order to satisfy present 
experimental bounds but still produce significant 
quantities detectable at other experiments.
It is not neccesary that the same mechanism be the 
responsible for all, the explanation of the absolute solar deficit, possible 
time variations of the signal 
and the production of antineutrinos. It is important then 
in this respect that the region of the parameter space where 
every effect is important be as large as possible.

Different concrete scenarios involving magnetic fields have been 
proposed where antineutrino generation is sizeable \cite{akh6,fio1}. 
The main problems of "hybrid" models where a spin-flavour magnetic 
moment transition gives $\nueL\to\anumR$ and mass-matter oscillations yields 
$\anumR\to\anueR$ transitions are to our opinion: 
a) The, nearly complete, lack of knowledge of the magnetic properties  in 
the interior of the Sun.
b) The high values, orders of magnitude higher than predicted by SM, of the 
magnetic moment which are needed to obtain any kind of sizeable effect.
c) The undesirable large time variations of the neutrino flux which 
 are usually associated with  them.

In this work we present a new mechanism for antineutrino production inside
the Sun which is largely  model independent.
We will consider the influence of a thin wall of 
highly chaotic  magnetic field on the propagation of neutrinos in matter.
For simplicity and in order to clearly expose the 
specific properties of the model we will disregard the influence of 
any other regular magnetic field.
A similar model, which includes an additional constant field but disregards
neutrino flavour mixing has been studied in \cite{tor3}.
As it has been noted in  \cite{tor3}, some advantages of neutrino 
propagation in  random magnetic fields include: 
a) The no neccesity to
know in detail specific magnetic profiles. b) The potentially stronger effect of the 
random magnetic field in comparison to regular magnetic field: for the latter 
the 
spin precession is roughly proportional to 
the quantity $\exp C\int dr B(r)$ where the integral is 
extended over the region where the 
magnetic field $B$ is present. However for the case of random magnetic fields
the precession is proportional to $\exp C' \int dr B^2(r)$. In addition the 
chaotic precession is irreversible and regeneration effects of the 
original neutrino flux are absent or reduced. 
c) It is still important to assume a large transition 
magnetic moment $\mu$ for the neutrino, but the dependence on it is substantially altered. While for regular oscillations  
the constant  $C\sim\mu$, for random magnetic fields $C'\sim \Omega^2 L_0 \mu^2$. The dependence on the magnetic moment $\mu$ can be traded off, at least partially, for the dependence on the parameters $\Omega^2,L_0$. These parameters describe the randomness and the coherence length of the magnetic field, they can be potentially very large, with an accepted range of variation of $\sim1-3$ orders of magnitude.
d) The overall influence of the random nature of the magnetic field is the 
flattening of the neutrino spectrum. 
The strong energy dependence on  resonances is reduced, so this scenario is a natural way to avoid strong time variations.

The plan for the rest of this work is as follows. In the next paragraphs we will justify qualitatively the presence of 
chaotic magnetic fields in the convective region of the sun; next we will 
summarize the formalism of the neutrino propagation in random media;
finally we will present the results. 
We will compute the antineutrino yield and total neutrino rate 
in Superkamiokande for different 
energy thresholds and compare with existing  bounds. 
The SNO antineutrino production should follow a similar trend as that one 
calculated here of SK.

\vspace{0.2cm}{\bf 3.} 
Very little is known, both theoretically and observationally,
 about the structure of the magnetic field inside the Sun 
\cite{park1,park2,vai1,vai2}.
The absence of a significant external poloidal field precludes the 
existence of an internal core field or at least strongly
limits  its magnitude. 
The evidence of surface magnetic activity and the solar cycle indicate 
however the existence of a certain degree of magnetic 
activity in the upper layer, the convective zone. 
The existence of surface magnetic fields
 spatially localized (the typical size of a sunspot is $\sim 1000$ km) 
which fluctuate  rapidly on time  has been shown to exist (\cite{ste2}, 
as cited in Ref.\cite{nic1}). 
It has been deduced  that a similar picture should hold in the convective zone.
On the other hand, the migration of the toroidal field towards the equator, the 11-year oscillation in sunspot number and the Hale-Nicholson law, can 
all be understood if it is assumed that  a dynamo mechanism is effective 
in the convective region of the Sun.
Direct numerical simulations of the  magnetohydrodinamic turbulence show that 
for a large magnetic  Reynolds number the magnetic field becomes indeed 
spatially intermittent \cite{vai2}.

It is generally  accepted  or at least strongly favored that a highly 
chaotic field, fluctuations of large amplitude and short length
scale ($L_0\approx 10^1-10^4$ Km), could be effective in the convective zone
 specially in the hydrodinamically unstable region separating it from the core. From   solar neutrino evidence: Kamiokande has shown  
repeteadly that there is not any significant semiannual variation of the neutrino flux. As it was pointed out in \cite{nic1,semi2}, if a magnetic interaction of $\nu_e$'s in the sun takes place, the 
Kamiokande result ''implies''  the dominance of the stochastic magnetic field over the large scale magnetic field.

In this work we will adopt the simplest option, we will ignore any large 
scale  magnetic fields and assume the existence of a thin region where a 
purely chaotic field is present.
Outside this layer a neutrino created at the solar interior experiences 
standard matter oscillations and the MSW effect.
While traversing the chaotic region, the master Equation 
governing the time evolution of the 4x4 (2 flavors are assumed by simplicity) 
 density matrix $\rho(t)$ is of the form:
\begin{eqnarray}
i\partial_t \rho&=&[H_{0},\rho]+\mu \tilde{B}_x(t)[V_x,\rho]+\mu \tilde{B}_y(t)[V_y,\rho].
\label{e7702}
\end{eqnarray}
The $\tilde{B}_x,\tilde{B}_y$ are the Cartesian transversal components
of the chaotic magnetic field. Vacuum mixing terms and matter terms corresponding to the SSM density profile 
are all included in $H_{0}$. 
The matrices $V_{x,y}$ are given in terms of the 
Pauli matrices $\sigma_{1,2}$ by:
\begin{equation}
V_x=\pmatrix{0&i \sigma_2\cr -i \sigma_2 & 0}, \quad
V_y=\pmatrix{0&-i \sigma_1\cr  i\sigma_1 & 0}.
\end{equation}

We assume that the components 
 $\tilde{B}_x,\tilde{B}_y$ are statistically independent, each of them 
characterized by a $\delta$-correlation function:
\begin{eqnarray}
\langle \tilde{B}_{x,y}(t)\tilde{B}_{x,y}(t')\rangle &=& \langle \tilde {B}_{x,y}^2\rangle  L_0\delta(t - t^{'}), \quad
\langle \tilde{B}_{x}(t)\tilde{B}_{y}(t')\rangle = 0 .
\end{eqnarray}
We will assume equipartition among components: 
$\langle\tilde{B}_x^2\rangle=\langle\tilde{B}_y^2\rangle=\langle\tilde{B}^2\rangle/3.$
The length scale $L_0$ is a basically unknown parameter.
It has been shown numerically and justified 
qualitatively in \cite{tor2} that the $\delta$-correlation function
 is a sufficiently good approximation to more realistic finite correlators
 even for relative large correlation lengths.

The averaged evolution equation is a simple generalization (see Ref. \cite{tor2} 
for a complete derivation) of the well known Redfield equation for two independent sources of noise \cite{lor1} and  reads 
($\Omega^2\equiv  L_0\mu^2 \langle\tilde{B}^2\rangle/3$):
\begin{eqnarray}
i\partial_t \avrho=[H_{0},\avrho]-i \Omega^2[V_x,[V_x,\avrho]]-i\Omega^2 
[V_y,[V_y,\avrho]].
\label{e8690}
\end{eqnarray}

It is possible to write the Eq.(\ref{e8690}) in a more evolved form. Taking into account  the particular form of the matrices $V_{x,y}$ 
and   rescaling  the density matrix according to: 
$$\avrhot=\exp(-4 \Omega^2  t)\avrhopt,$$
 the double commutators simplify and the evolution equation finally reads:
\begin{eqnarray}
i\partial_t \avrhop=[H_{0},\avrhop]+i 2\Omega^2 \left (V_x\avrhop V_x+V_y\avrhop V_y\right ).
\label{e8692}
\end{eqnarray}

It is useful however to consider the 
solution to  Eq.(\ref{e8692})  when $H_{0}\equiv0$. 
This is the appropriate limit when dealing with extremely low $\Delta m^2$ or 
very large energies, for an extreme level of noise or when the 
distance over which the noise is acting is small enough to consider the 
evolution driven by $H_{0}$ negligible.
 In any other scenario it can give at least an idea of the general
behavior of the solutions to the full  Eq.(\ref{e8692}). 
When $H_0=0$   only the two last terms   in the equation  remain 
and an
exact simple expression is obtainable by ordinary algebraic methods.
The full 4x4 Hamiltonian decouples in 2x2 blocks.
The quantities of interest, the averaged transition probabilities,
 are given by the diagonal elements of $\avrho$. 
If $P_{f,i}$ are the final  and initial probabilities 
( at the exit and at the entrance of  the noise  region) 
their   averaged counterparts  fulfill 
linear relations among them, schematically:
\begin{eqnarray}
Q_f^{A,B}&=&M Q_i^{A,B}
\label{e4501}
\end{eqnarray}
with $Q^{A,B}$ any of the two dimensional vectors
\begin{equation}
Q^A=\pmatrix{\langle P(\nu_{eL}\to  \nu_{eL}  )\rangle \cr 
\langle P(\nu_{eL}\to \tilde{\nu}_{\mu R} )\rangle }; \quad
Q^B=\pmatrix{\langle P(\nu_{eL}\to  \tilde{\nu}_{eR}  )\rangle \cr 
\langle P(\nu_{eL}\to \nu_{\mu L} )\rangle } \quad
\label{e4502}
\end{equation}
and the Markovian matrix $M$ (with 
 $P=1/2\left (1+ \exp(-4 \Omega^2 \Delta t)\right)$):
\begin{eqnarray}
M&=&\pmatrix{P & 1-P \cr 1-P & P}.
\end{eqnarray}
The quantity $P$   is a good approximation for the 
 depolarization that the presence of noise induces in the averaged 
neutrino density matrix.  $\Delta t$ is the distance over which 
the noise is acting.

From Eqs.(\ref{e4501}-\ref{e4502}) we have in particular the relation
\begin{equation}
\langle P(\nu_{eL}\to  \tilde{\nu}_{eR}  )\rangle_f  = (1-P) 
\langle P(\nu_{eL}\to \nu_{\mu L} )\rangle_i
\end{equation}
in the case of the initial number of electron 
antineutrinos at 
the entrance of the noise wall being zero. 
The final number is  proportional to the initial number of 
muon neutrinos. 
This number might be large if  $1-P$ is not small and if
 the neutrino have passed through a MSW resonance  before arriving to 
the noise region. The MSW resonance 
 converts practically all the initial $\overline{\nu}_e$ flux 
into $\overline{\nu}_{\mu}$. 
The fluctuating magnetic field converts them into 
$\overline{\nu}_e$. Part of the electron 
antineutrinos will be reconverted into  $\overline{\nu}_{\mu}$ by mass 
oscillations (vacuum and non-resonant matter oscillations). This reconversion
is limited in this case, in contrast with 
regular magnetic fields, by the irreversible character of  Eq.(\ref{e8692})

\vspace{0.2cm}{\bf 4}. 
The averaged master equation (\ref{e8692}) has been integrated numerically 
for a variety of mixing parameters ($\Delta m^2, \sin^2 2\theta$) and 
randomness parameter $P$.
The parameter $P$ has been varied between its maximum and minimum 
values (respectively 1, absence of magnetic field, 1/2, complete depolarization
of the density matrix). The corresponding  r.m.s fields  are
in the range  $\surd{\langle \tilde{B}^2\rangle}=0- 600 $ kG 
(supposing the scale $L_0=1000$ Km and $\mu=10^{-11}\mu_B$).

We have defined the weighted electron antineutrino appearance probability:
\begin{eqnarray}
\langle P_{\nu_e\overline{\nu}_e}\rangle
&=&\frac{\int_{E>Eth} dE \ \sigma(E) \Phi(E) P_{\nu_e,\overline{\nu}_e}(E)}
{\int_{E>Eth} dE\  \sigma(E) \Phi(E)} 
\label{e4545}
\end{eqnarray}
where $\sigma(E),\Phi(E)$ are respectively the differential cross section for the 
isotropic background process 
$\overline{\nu}_e+p\rightarrow e^+ n$ and the differential total neutrino flux coming from the Sun according to
the SSM \cite{BP95}. 
A constant experimental detector efficiency has been assumed over all the 
energy range.
Three different illustrative  threshold energies have been 
chosen $E_{th}=8.3,5.5$ and $2.0$ MeV.
For $E_{th}<7.0$ MeV, the weighted probability varies modestly  because
the product $\sigma\time\Phi$ is strongly peaked around $E\approx 9-10$ MeV.

\begin{figure}[p]
\centering\hspace{0.8cm}
\epsfig{file=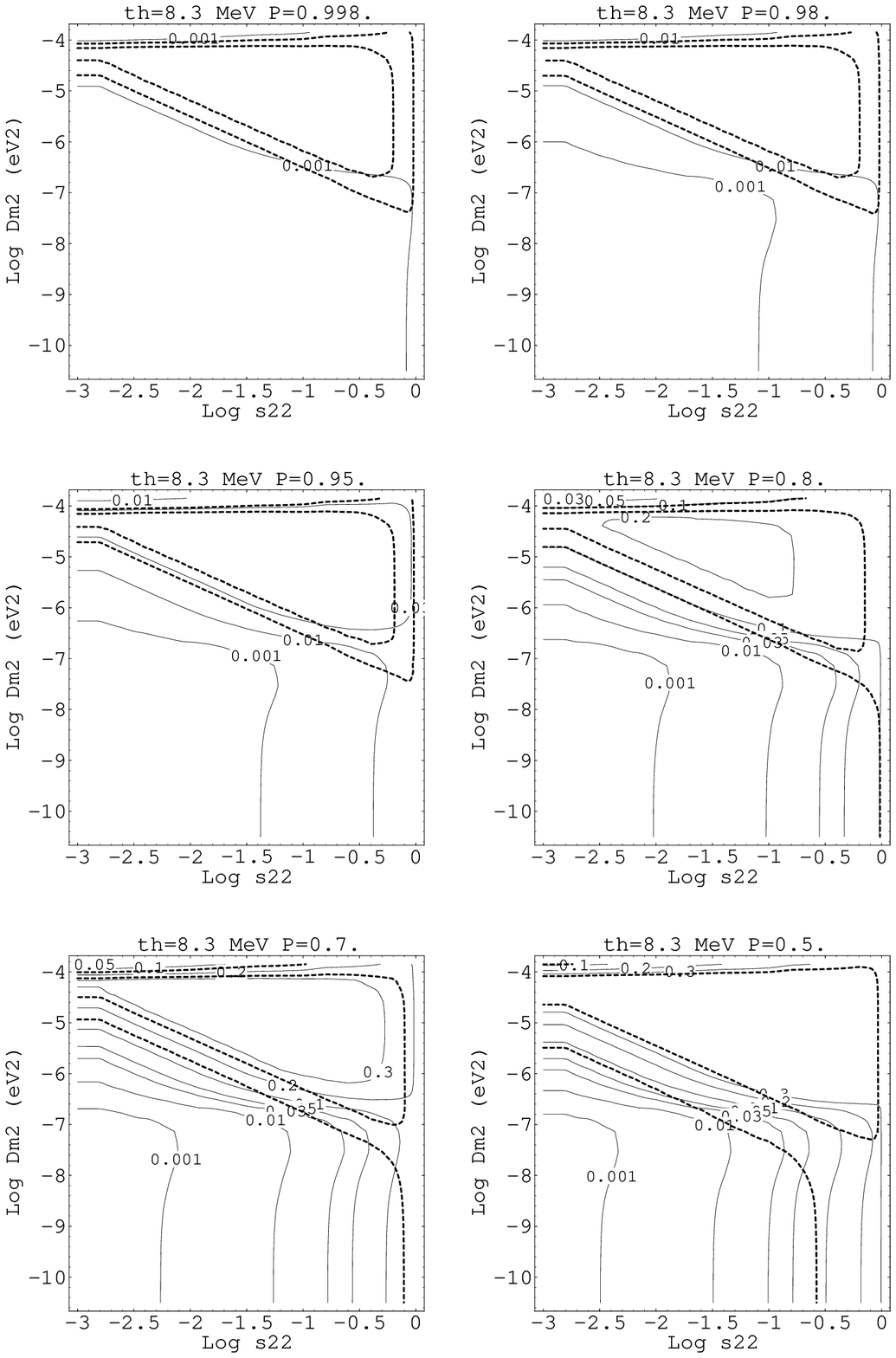,height=16cm}
\caption{
Continuous lines:
$\protect\overline{\nu_e}$ production as a function of neutrino mixing angle 
$\log \sin^2 2\theta$ and $\Delta m^2$ (eV$^2$). The average probability, Eq.(\protect\ref{e4545}), is given for different levels of noise. 
Dashed thick lines (not labeled): $S/S_{SSM}$ rates at SK, contours 
correspond to $S/S_{SSM}=0.3,0.5$
The threshold 
neutrino energy is in this case: $E_{th}=8.3$ MeV.
From left to right and from top 
to bottom: $P=0.998,0.98,0.95,0.8,0.7,0.5$. 
The corresponding  r.m.s fields  are
 $\surd{\langle \tilde{B}^2\rangle}=15,45,65,150,220$ and 600 kG
respectively  (supposing the scale $L_0=1000$ Km and $\mu=10^{-11}\mu_B$).}
\label{f20a830Nall}
\end{figure}

\begin{figure}[p]
\centering\hspace{0.8cm}
\epsfig{file=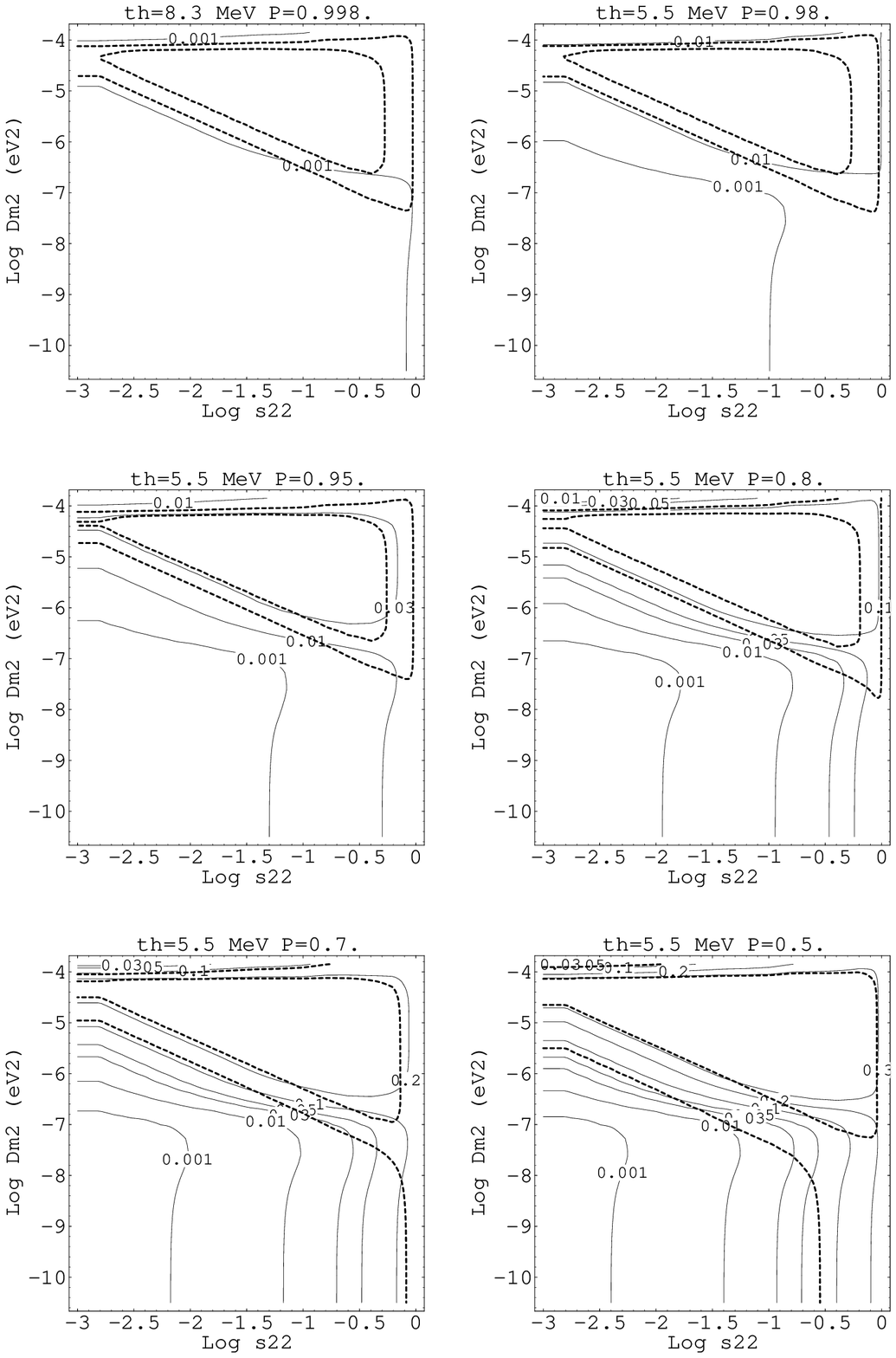,height=16cm}
\caption{ Idem as Fig.(\protect\ref{f20a830Nall}) for  $E_{th}=5.5$ MeV.
The dashed line corresponds to the existing experimental bound 
($P<5 \%$) from SK. }
\label{f20a500Nall}
\end{figure}

\begin{figure}[p]
\centering\hspace{0.8cm}
\epsfig{file=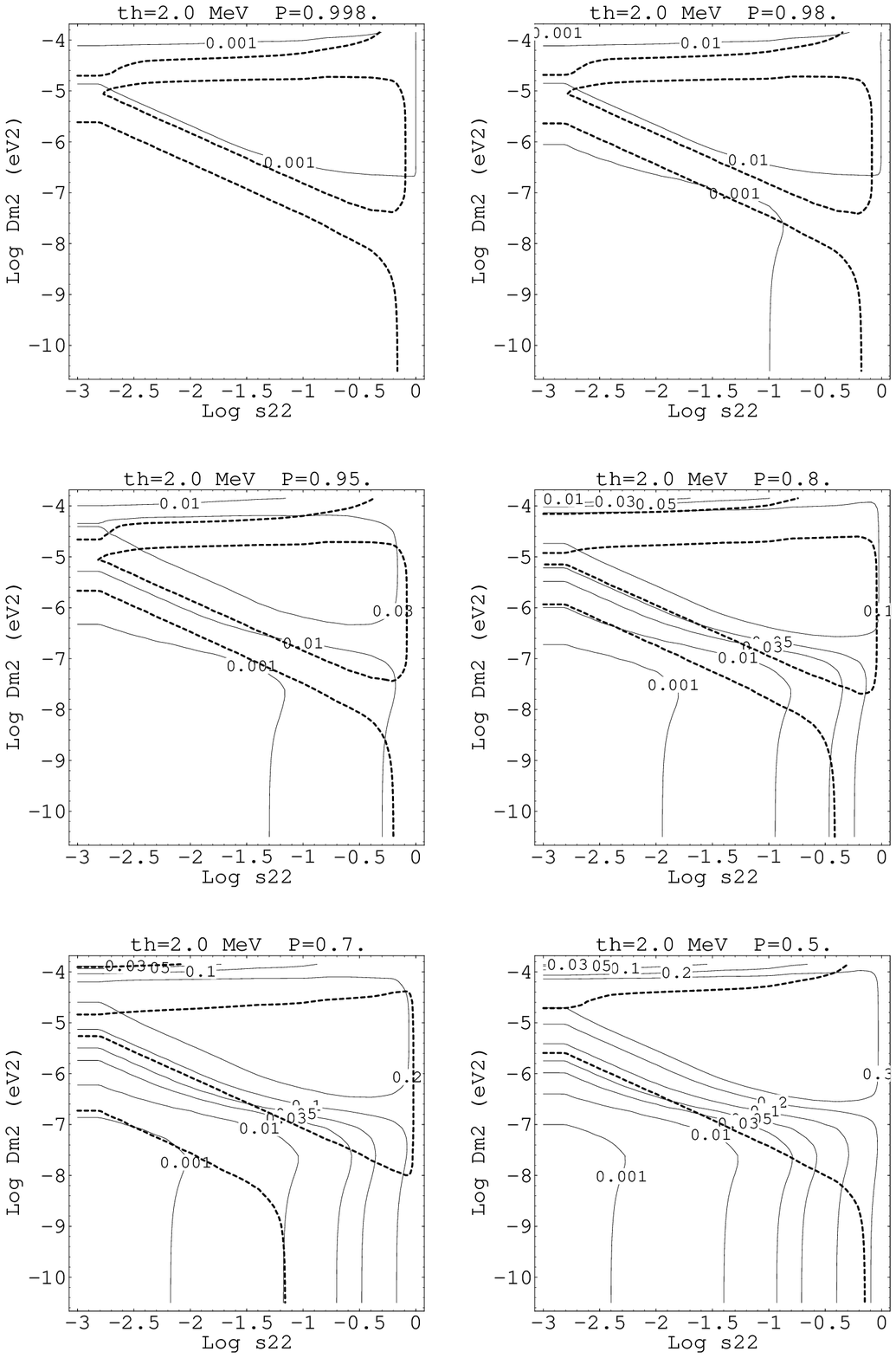,height=16cm}
\caption{ Idem as Figs.(\protect\ref{f20a830Nall})
and (\protect\ref{f20a500Nall})
 for an energy threshold $E_{th}=2.0$ MeV.}
\label{f20a100Nall}
\end{figure}

In calculating the expected total signal rates, 
the elastic cross sections for 
the processes $\nu_x e\to \nu_x e$ with 
$\nu_x=\nu_e,\nu_\mu,\overline{\nu}_e,\overline{\nu}_\mu$ 
has been taken from  \cite{pas1}. 
The results are shown  
in Figs.(\ref{f20a830Nall},\ref{f20a500Nall},\ref{f20a100Nall}) 
corresponding  respectively to each of the threshold energies
which are considered.
The continuous lines in the figures correspond to the contours for 
$\langle P_{\nu_e\overline{\nu}_e}\rangle$. The dashed thick lines correspond 
to the ratio of theoretical to SSM predicted total signal (S/S$_{SSM}$).

 The characteristic MSW triangular contours  are apparent in the 
figures, coinciding for both  antineutrino and total rates.
Note however how they shift independently  as the threshold energy 
decreases.
As the noise level increases,  the 0.5 contour 
expands and the region  compatible with present SK data enlarges. 
When combining  total rates for the different solar  neutrino experiments, 
the global  allowed regions appear  roughly at the  three corners of 
the triangular regions (regions labeled as SMA, LMA and LOW MSW in, 
for example, Ref. \cite{bah2}).

When the noise level is low ($P=0.95$ or higher), 
the bound $\langle P_{\nu_e\overline{\nu}_e}\rangle<\approx 0.05$ 
 obtained \cite{akh6,fio1} from Kamiokande data is fulfilled for any 
point of the $(\Delta m^2,\sin^2 2\theta)$ parameter space considered. 
Nevertheless the antineutrino yield is non-negligible and is situated, 
in the range of values which could be experimentally tested 
 in the near future \cite{fio1}. 
For $P\approx 0.98$ there is a significant antineutrino
 production (1-3\%) 
for practically all  the region 
compatible with the total signal rate. 
This affirmation would continue to be true if the rest of solar 
neutrino experiments are
taken into account (LMA and SMA regions).
Only for the  lowest $\Delta m^2$ compatible with SK total rate (the 
would-be LOW MSW region) we obtain an antineutrino probability smaller than
1\%. 
Similar comments can be extracted from the plot corresponding to
 $P=0.95$. In this case the antineutrino yield increases to the 
$3-5\%$ level for the LMA and SMA regions and to $1-3\%$ for the 
LOW MSW region.
The predictions  for the $5.5$ and $2 $ MeV thresholds 
are of similar character and no fundamental changes are observed: Similar levels of antineutrino production are expected in regions compatible
with higher energy bounds.

Values for the noise parameter of 
$P=0.8$ or  lower are not favored when we consider the combined 
total rates from the existing solar neutrino experiments at least in the
limit of negligible mixing angle (see Ref.\cite{tor3}).  
Let's note that for these levels of noise, 
regions with a very high antineutrino yield are possible 
($\langle P_{\nu_e\overline{\nu}_e}\rangle\approx 20-30\%$). 
Simultaneously, the
 region with a total rate compatible with SK results enlarges in the large angle limit. However  the
antineutrino production serves as a strong constraint: the 
consideration of the SK total 
rate and antineutrino production 
 make  that  only marginal areas  with 
$\Delta m^2\approx 10^{-8}$ eV$^2$  are still allowed. 
The same situation repeats at the $5.5$ MeV threshold. 
At $E_{th}=2$ MeV 
the predicted yield  increases  but is  always in the $\sim<3\%$ level if 
we restrict ourselves to the areas allowed at higher 
thresholds.

\vspace{0.2cm}{\bf 5.}  In conclusion,
in this work we have considered a model where solar neutrinos pass 
through a thin layer where chaotic magnetic field is present.
Some comment is in order about the nature of the chaotic field 
considered here.
According to  \cite{tor2}  the $\delta$-correlation function
 is a sufficiently good approximation to more realistic finite correlators
 even for a relative large correlation length.
Moreover, the quantity $P$ is  connected with the depolarization 
of the density matrix, and therefore,
it  offers some advantage 
with respect to more ''physical'' parameters 
 as $\Omega^2$ or
$\surd \langle\tilde{B}^2\rangle$: we  expect that  results 
expressed in terms of $P$ are largely independent of the 
concrete modelizing of  the stochastic properties of 
the fluctuating field (in particular the form of the correlator).

From the numerical results presented above, it is deduced that a  small but significant quantity of antineutrinos is expected to be detectable 
in SuperKamiokande and 
other water Cerenkov experiments (SNO) in parameter regions compatible with present 
experimental bounds (including neutrino  magnetic  moment bounds).
If the neutrino is a Majorana particle, 
the scenario proposed  in this work predicts a 
significant quantity of antineutrinos (1-3\%) even for a modest 
level of noise ($P>0.95$) in regions compatible with existing experimental evidence. Lowering 
the detection energy threshold the quantity of antineutrinos tends to increase, but only very gently.
In the other hand it would be  possible to obtain information about the 
solar magnetic internal field if 
antineutrino bounds reach the $1\%$ level and a particle physics 
solution to the SNP is assumed.  
If a  squared mass
 difference  of the order $\dms\approx 10^{-5}$ is preferred,
there are two favored alternatives, either the neutrino is not a Majorana
 particle or the level of solar convective magnetic noise is 
small ($P\approx 1$). 
This is so  because otherwise we would have too many antineutrinos at 
high energy. Note however that even in the most restrictive case 
  a significant quantity of antineutrinos ($\sim 1\%$)
might be detectable 
at lower energies close to the $n$ production threshold.

\vspace{2cm}

{\bf Acknowledgments}. 

I acknowledge V.B.  Semikoz for drawing my attention to the 
problem, for many and useful discussions 
about solar magnetic fields and for pointing out to me numerous references.
This work has been supported by DGICYT under Grant 
 PB95-1077 and by  a DGICYT-MEC contract  at Univ. de Valencia.

\newpage 

\end{document}